\documentclass[english,prb,twocolumn]{revtex4}

\usepackage{graphicx}
\usepackage{mathrsfs}
\usepackage{float}
\usepackage{dcolumn}
\usepackage{bm}
\usepackage{epsfig}
\newcommand{\ignore}[1]{}

\begin{document}

\title[]{A first principles study on FeAs single layers}

\author{Jun Dai, Zhen-yu Li, Jin-long Yang\footnote{
Corresponding author. E-mail: jlyang@ustc.edu.cn} }

\address{Hefei National Laboratory for Physical Sciences at
Microscale, University of Science and Technology of China, Hefei,
Anhui 230026, P.R. China}

\date{\today}

\begin{abstract}
FeAs$^-$ single layer is tested as a simple model for LaFeAsO and
BaFe$_2$As$_2$ based on first-principles calculations using
generalized gradient approximation (GGA) and GGA+$U$. The calculated
single-layer geometric and electronic structures are inconsistent
with that of bulk materials. The bulk collinear antiferromagnetic
ground state is failed to be obtained in the FeAs$^-$ single layer.
The monotonous behavior of the Fe-As distance in $z$ direction upon
electron or hole doping is also in contrast with bulk materials. Our
results indicate that, in LaFeAsO and BaFe$_2$As$_2$, interactions
between FeAs layer and other layers beyond simple charge doping are
important, and a single FeAs layer may not represent a good model
for Fe based superconducting materials.

\end{abstract}

\maketitle

\section{INTRODUCTION}
The recently discovered high temperature superconductivity in
LaFeAs[O$_{1-x}$F$_x$]\cite{jacs} has attracted a lot of interests
and triggered the research for other iron-based superconductors. Up
to now, most of the Fe-based superconductors are based on either
$R$FeAsO ($R$=La, Ce, Sm, Nd, Pr and Gd) or $A$Fe$_2$As$_2$ ($A$=Ba,
Sr) structures. In the former 1111 series, replacing La with other
rare-earth atom increases the transition temperature (T$_c$) up to
55 K for SmO$_{1-x}$F$_x$FeAs. \cite{Sm1, Sm2} For the latter 122
series, with appropriate alkali metal (K and Cs) doping, T$_c$ can
be raised up to 37 K.\cite{122-1,122-2} Very recently,
superconductivity has also been reported for As-free material
FeSe$_{1-\delta}$ and Fe(Te$_x$Se$_{1-x}$)$_{1-\delta}$ with T$_c$
around 27 K under pressure.\cite{FeSe1,FeSe2} Moreover, replacing
$R$O layer in 1111 materials with Li or Na also leads to T$_c$ of 18
K and 9 K respectively.\cite{Li, Na}

\begin{figure}[!hbp]
\includegraphics[width=7.5cm]{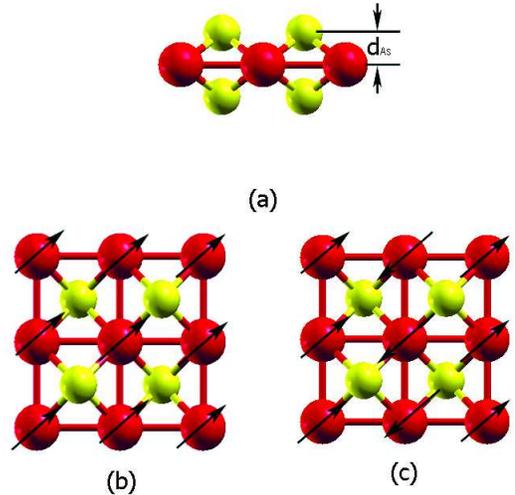}
\caption{(Color online) The crystal structure of the
$\sqrt{2}\times\sqrt{2}\times1$ single FeAs layer. The red and
yellow balls are Fe and As species respectively. (a) is the side
view of single FeAs layer, (b) and (c)  are the top view of FeAs
layer, where the black arrows signify the FM (b) and collinear AFM
(c) configurations on Fe atoms. } \label{fig1}
\end{figure}

The Fe-based superconductors have a quasi two-dimensional
tetrahedral structure, where FeAs layers are separated by $R$O
($R$=La, Ce, Sm, Nd, Pr and Gd), $A$ ($A$=Ba, Sr), or Li (Na)
layers. Except LiFeAs and NaFeAs, both parent compounds of 1111 and
122 superconductors are metallic but not superconducting. They
undergo a phase transition from tetragonal to orthorhombic with the
decrease of temperature, which accompanies with a new collinear
antiferromagnetic (AFM) order, known the SDW phase. \cite{AFM1,
AFM2, AFM3, AFM4} Upon doping the SDW is suppressed, and
superconductivity appears. First principles calculations for both
1111 and 122 materials have been reported, using either local spin
density (LSDA) or generalized gradient approximations (GGA).
\cite{theo1, theo2, theo3, theo4, theo5, theo6} The collinear AFM
ground state in the parent compounds has been confirmed by theory.

Although the mechanism of superconductivity in these Fe-based
materials is still unknown, it is clear the essential physics lies
in the common FeAs layer. It serves as the conducting layer, and the
interplay between magnetism and superconductivity happens in this
layer. Therefore, it is important to investigate its structure
change as well as the evolution of electronic properties upon
doping.

At low temperature (25 K), in F doped LaFeAsO , with F doping, the
Fe-As bond length changes less than 0.1\%, while the La-As distance
reduces by $\sim$1.5\% and the La-O distance increases by
$\sim$0.8\%. \cite{sst21125028} These results demonstrate that the
structure of FeAs layer changes slightly upon doping, in contrast
with the significantly modified LaO layer. Besides, the layered
structure of Fe-based superconductors is very similar to that of
cuprates and Na$_x$CoO$_2$ superconductors. A single CoO$_2$ layer
has been successfully used as a model system to investigate the
doping effects on Na$_x$CoO$_2$. \cite{CoO1,CoO2} Thus, one question
comes out, can we use a similar model of single FeAs layer to study
the doping effects on Fe-based superconductors?

In this article, we calculate the geometric, electronic, and
magnetic properties of a singe FeAs$^-$ layer in the framework of
density functional theory (DFT).  We fail to obtain the collinear
AFM phase with both optimized structure and experimental structure.
The behavior of $d_{As}$ (Fig. \ref{fig1}a) respecting to the doping
level also differs from that of 1111 and 211 materials.

\section{MODEL AND METHOD}

As shown in Fig.\ref{fig1}, FeAs layers are formed by edge-shared
FeAs$_4$ tetrahedras with Fe ions sandwiched between two As sheets.
In the undoped parent materials for both the 1111 and 122 series,
the FeAs layer is negatively charged with one electron, which means
undoped FeAs layer corresponds to FeAs$^-$. Starting from this
point, hole or electron doping is realized by simply adding or
removing electrons from FeAs layer, with uniform compensated charge
background. Doping level $x$ is defined for charged system
FeAs$^{-(1+x)}$. In this work, doping level $x$ ranging from $-1.00$
to $+1.00$ is investigated.

Most of the theoretical works on the electronic and magnetic
properties of Fe-based superconductors in the literature are based
on the high-temperature tetrahedral structure, with $a$=$b$.
\cite{theo1, theo2, theo3, theo5, As1} One attempt to obtain the
low-temperature orthorhombic lattice constants used an alternative
strategy with the magnetic moments fixed to experimental value
instead of optimizing on ground state potential energy surface.
\cite{theo4} For single layer in our case, if we directly scan the
orthorhombic structure parameters, we obtain $a$ and $b$ around 4.64
\AA, which is significantly lower than the experimental values (by
more than 1.0\AA). Therefore, in the following calculations, we fix
$a$ and $b$ to their experimental values 5.683 and 5.710,
respectively. \cite{sst21125028} In order to exclude interactions
between neighboring layers, $c$ is set to 14 \AA, corresponding to
about 10 \AA's distance between two neighboring FeAs layers. The
positions of all atoms are allowed to relax until forces on each
atoms are smaller than 0.01 eV/\AA. Due to the symmetry of the
system, the only degree of freedom of the atoms is the $z$
coordinate of As. As shown in Fig. \ref{fig1}a, $z$ coordinate of As
related to the Fe plane is marked as $d_{As}$.

The electronic structure calculations are carried out using the
Vienna \textit{ab initio} simulation package. \cite{vasp} PBE
functional is used.\cite{GGA} The electron-ion interactions are
described in the framework of the projected augment waves method and
the frozen core approximation.\cite{PAW} The energy cutoff is set to
be 600 eV, the same as previously used for LaOFeP.
\cite{PRB-Lebegue} For density of states (DOS) calculation, we used
a 12$\times$12$\times$6 Monkhorst-Pack $k$-point grid to sample the
Brillouin zone, while for geometry optimization, a
8$\times$8$\times$4 grid have been used.

For magnetic property calculations, initial magnetic moments are set
according to non-spin polarized (NM), ferromagnetic (FM), and
collinear antiferromagnetic (AFM) ordering. The FM and collinear AFM
configurations are illustrated in Fig.\ref{fig1} b and c. The latter
is the experimentally observed ground state for parent compounds.

In GGA+U calculations, we adopt a simplified model, where the
on-site Coulomb repulsion $U$ and the atomic-orbital intra-exchange
energy $J$ are simplified to one parameter $U_{eff}=U-J$. For
simplicity, we will call $U_{eff}$ as $U$ hereafter.

\begin{figure}
\includegraphics[width=7.5cm]{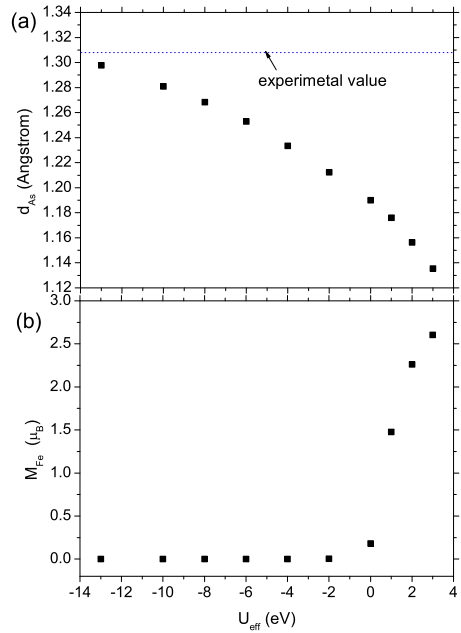}
\caption{ Dependence of (a) $d_{As}$ and (b) magnetic moment
($\mu_B$/Fe) on Hubbard parameter U. The experimental value of
$d_{As}$ for LaFeAsO is marked by a dotted line. } \label{fig2}
\end{figure}

\section{RESULTS AND DISCUSSIONS}

First we performed structure optimization and electronic structure
calculations for FeAs$^-$. From GGA results, the total energy of FM
state is about 0.002 eV lower than that of NM state, and the
magnetic moment on Fe in FM state is about 0.18 $\mu_B$. The
collinear AFM state, which is reported to be the ground state of
1111 and 122 materials, is not stable at all.

The optimized $d_{As}$ is 1.19 \AA, which is significantly lower
than the experimental value for LaFeAsO (about 1.31 \AA\ at 4
K).\cite{AFM1} The electronic structure as well as the magnetic
moments on Fe in 1111 and 122 parent compounds are very sensitive to
$d_{As}$, \cite{As1,As2} so the discrepancy between the calculated
magnetic structure and the experimental one may cause by the heavily
underestimated $d_{As}$.

Considering the possible electron correlation in FeAs, GGA+U may be
a useful strategy to correct $d_{As}$. Different U values are used
to test it effects. In all DFT+U calculations, we get either an FM
or a NM ground state. The magnetic moment increase with $U$ for
medium value of $U$, and when $U$ goes to relative large negative
value, the magnetic moment on Fe will be quenched, which is
consistent with the trend that negative $U$ delocalize electron. As
shown in Fig.\ref{fig2}, $d_{As}$ decreases monotonously with $U$.
To get the experimental $d_{As}$ value, an unphysical $U$ as low as
-13.0 eV should be used.

\begin{figure}
\includegraphics[width=7.5cm]{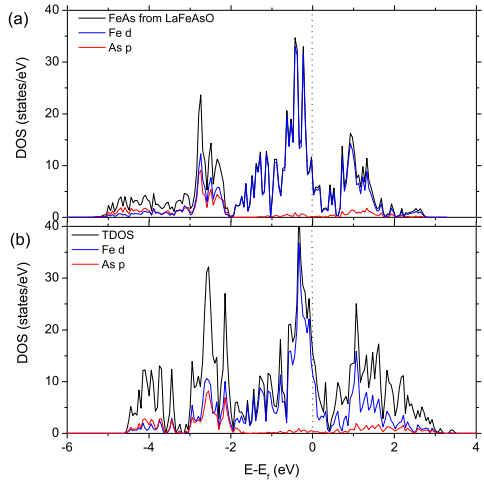}
\caption{(Color online) Density of states of FeAs and the
contributions of Fe d states and As p states to FeAs DOS. (a) PDOS
of FeAs in LaFeAsO, (b) DOS of FeAs single layer with experimental
values of Fe and As positions. The Fermi level is aligned to 0.00
eV. } \label{fig3}
\end{figure}

Since DFT+U also fail to give the correct $d_{As}$, another thing we
can try is to directly adopt the experimental $d_{As}$ to see if we
can get correct magnetic properties. The calculated total energy of
FM state is about 0.0278 eV/cell lower than the NM state, with a
magnetic moment of 0.48 $\mu_B$ per Fe atom. The collinear AFM state
does not exist again in single FeAs layer with experimental
$d_{As}$. The resulting electronic density of states (DOS) of
undoped FeAs layer is shown in Fig.\ref{fig3}, comparing to the FeAs
partial DOS (PDOS) of LaFeAsO obtained with experimental structure
parameters at low temperature. \cite{sst21125028,AFM1} The main
structures of the DOS and PDOS are very similar. In energy ranges
from -2 eV to 2 eV near the Fermi level, Fe 3d states dominate. The
peaks from -3 to -2 eV are mixed Fe and As states. The density of
states at Fermi level ($N(E_{f})$) for single FeAs layer is
significantly larger than that of FeAs PDOS of LaFeAsO ($\sim$16.13
eV$^{-1}$ v.s. $\sim$7.15 eV$^{-1}$), which may result in different
magnetic behavior.

\begin{figure}
\includegraphics[width=7.5cm]{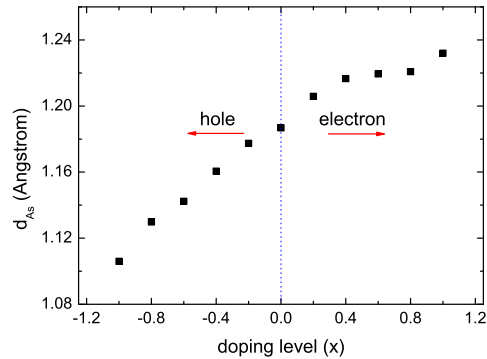}
\caption{(Color online) The evolution of $d_{As}$ with the doping
level $x$. } \label{fig4}
\end{figure}

The doping effect for single FeAs layer is also studied.
Experimentally, the structure transition from tetrahedral phase to
orthorhombic phase is suppressed when sufficient electron or hole
doping is applied, and the $d_{As}$ in tetrahedral and orthorhombic
phases in undoped LaFeAsO is almost the same. \cite{sst21125028}
Therefore, the optimization for FeAs single layer with doping level
from -1.0 to 1.0 is carried out using the experimental tetrahedral
phase lattice parameters (a=b=5.706 \AA). As shown in Fig.
\ref{fig4}, the $d_{As}$ changes monotonously with doping level, to
be specific, increases with the level of electron doping, and
decreases with the level of hole doping. In experiment, at 120 K for
LaFeAsO, $d_{As}$ do increase slightly from 1.319 to 1.323 \AA\,
when 14\% F is doped. \cite{sst21125028,struc} However, in the case
of hole doping, for Ba$_{1-x}$K$_x$Fe$_2$As$_2$ at 10 K, $d_{As}$
increases from 1.344, to 1.351 \AA\ and 1.358 \AA\, when hole doping
increases from zero to 10\% and 20\%, respectively. \cite{AFM2,Ba}
So the calculated trend of $d_{As}$ using the single layer FeAs
model is insufficient to describe the doping effects on geometrical
properties of Fe-based superconductors.

\section{Conclusion}
We have performed first-principles calculations on single layer
FeAs. In this model where the inter layer interaction is ignored, we
find the structure of the FeAs layer in $R$FeAsO and $A$Fe$_2$As$_2$
can not be reproduced accurately in the framework of GGA and
GGA+$U$. Besides, with both optimized and experimental lattice
parameters, the collinear AFM ground state of $R$FeAsO and
$A$Fe$_2$As$_2$ can not be obtained in FeAs single layer.

In the simple single layer model, the inter-layer interactions
between $Re$O ($R$) layers and FeAs layers are excluded. Our results
suggest that this interactions may need to be considered to obtain
correct geometry. This conclusion is important for choosing a proper
theoretical model in future investigation of Fe based
superconductors.

\section*{ACKNOWLEDGMENTS}
This work was partially supported by the National Natural Science
Foundation of China under Grant Nos. 50721091, 20533030,
50731160010, 20873129, and 20803071, by National Key Basic Research
Program under Grant No. 2006CB922004, by the USTC-HP HPC project,
and by the SCCAS and Shanghai Supercomputer Center.

\textbf{Note added after submission:} A recent experiment suggested
a nearly isotropic superconductivity in
(Ba,K)Fe$_2$As$_2$.\cite{iso}

\end{document}